# QuantumOptics.jl: A Julia framework for simulating open quantum systems

Sebastian Krämer[a], David Plankensteiner[a,*], Laurin Ostermann[a], Helmut Ritsch[a]

[a]*Institut für Theoretische Physik, Universität Innsbruck, Technikerstrasse 21, A-6020 Innsbruck, Austria*

**Abstract**

We present an open source computational framework geared towards the efficient numerical investigation of open quantum systems written in the Julia programming language. Built exclusively in Julia and based on standard quantum optics notation, the toolbox offers speed comparable to low-level statically typed languages, without compromising on the accessibility and code readability found in dynamic languages. After introducing the framework, we highlight its features and showcase implementations of generic quantum models. Finally, we compare its usability and performance to two well-established and widely used numerical quantum libraries.

*Keywords:* Quantum Optics; Quantum Mechanics; Numerics; Julia language;

**PROGRAM SUMMARY**

*Program Title:* QuantumOptics.jl
*Licensing provisions:* MIT
*Programming language:* Julia

*Nature of problem:* Dynamics of open quantum systems

*Solution method:* Numerically solving the Schrödinger or master equation or a Monte Carlo wave-function approach.

*Additional comments including Restrictions and Unusual features:*
The framework may be used for problems that fulfill the necessary conditions such that they can be described by a Schrödinger or master equation. Furthermore, the aim is to efficiently and easily simulate systems of moderate size rather than pushing the limits of what is possible numerically.

## 1. Introduction

Numerical simulations of open quantum systems are essential to research fields like quantum optics or quantum information as the number of analytically solvable systems is quite limited. Due to the usefulness of such numerical calculations to study systems and phenomena otherwise only accessible through elaborate experimental tests, it is of interest to make said numerical calculations as approachable as possible to a wide audience without compromising too much on their efficiency. In this form they can also be a useful tool in teaching. An early, greatly successful attempt in this direction has been the Quantum Optics (QO) Toolbox in Matlab [1], which dates back almost two decades. Other approaches [2, 3] have mostly focused on efficiency, thus sacrificing accessibility by employing lower level languages like C++ and template metaprogramming. None of these, however, have managed to gain a popularity comparable to the QO Toolbox.

It was not until recent years, that a toolbox similar to the one in Matlab called QuTiP (Quantum Toolbox in Python) has been developed [4, 5], which in some sense superseded the QO Toolbox. QuTiP's wide adaptation can be traced back to a couple of advantageous properties: QuTiP as well as its underlying language, Python, are both open source, a fact that is greatly appreciated in the scientific community. When compared to the QO Toolbox, which runs on the proprietary Matlab, QuTiP is equally convenient to use and switching to it requires very little effort. Additionally, due to its open development approach the project has acquired many active contributors since its debut and as a conse-quence it contains many features that go beyond QO Toolbox's capabilities. Another outstanding feature is QuTiP's extensive documentation and the considerable amount of available examples. One downside, however, is the fact, that any time-critical calculations need to be outsourced via Cython [6] or performed by external libraries written in e.g. C, C++ or Fortran. This not only applies to the framework itself, but is also a concern for any user. If some code provided by the user happens to be time-critical, he or she then has to port it to a low-level programming language.

*Corresponding author.
E-mail address:* david.plankensteiner@uibk.ac.at

The programming language Julia [7, 8] avoids this problem by offering a clean and convenient syntax typically associated with dynamic languages while at the same time providing speed comparable to compiled languages due to its just in time (JIT) compilation. For this reason, it has been gaining a lot of momentum in the community of scientific programming already, even though it is still under active development. While Python packages like PyPy [9] or Numba [10] will also allow for JIT compilation in the same fashion as Julia, with the later even relying on the same LLVM compiler, Python was designed to be an interpreted language and thus only a subset of functions will benefit from the JIT functionality. In contrast, Julia was created with the JIT paradigm in mind and no extra effort by the user is required.

Taking full advantage of its easy-to-read syntax and its efficiency, we built a new open source framework, *QuantumOptics.jl*, written entirely in Julia. It is specifically geared towards the efficient and easy numerical simulation of open quantum systems. In this paper, we demonstrate the capabilities of our toolbox in its current version v0.4.1. We show that it offers speed in numerical calculations, while at the same time the source code remains intuitive and easily accessible.

Our framework can be installed very straightforwardly: after having setup Julia itself, where detailed instructions can be found on the Julia website [11], one can make use of Julias's package manager and simply execute the following command:

```
Pkg.add("QuantumOptics")
```

To obtain a first impression of QuantumOptics.jl consider code sample 1, which simulates the well-known Jaynes-Cummings model. A two-level atom with a transition frequency $\omega_\mathrm{a}$, modeled as a spin-1/2 particle, coherently couples to a cavity mode of frequency $\omega_\mathrm{c}$. Initially, the particle is in the ground state and the field mode is prepared in a coherent state $|\alpha\rangle$. The time evolution of the system is governed by the Schrödinger equation

$$i\frac{\partial \psi}{\partial t} = H_\mathrm{JC}\psi, \quad (1)$$

where the Hamiltonian, given in a suitable reference frame reads ($\hbar = 1$)

$$H_\mathrm{JC} = \Delta a^\dagger a + g\left(a^\dagger \sigma^- + a\sigma^+\right). \quad (2)$$

Here, $\Delta = \omega_\mathrm{c} - \omega_\mathrm{a}$ and $g$ is the coupling strength between the atom and the cavity. Furthermore, $a$ ($a^\dagger$) is the photon annihilation (creation) operator of the cavity and $\sigma^-$ ($\sigma^+$) denotes the atomic dipole's lowering (raising) operator. As one can see from code sample 1, the framework predefines all the necessary operators, which enables one to implement the above model in a few simple lines of code.

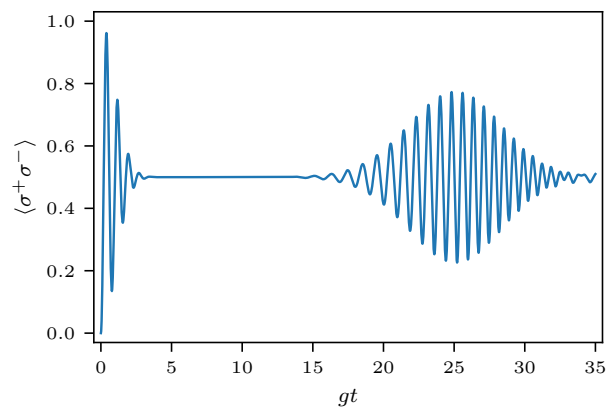

Figure 1: *Jaynes-Cummings model: atomic population dynamics.* The plot was created with the results from code sample 1.

```
using QuantumOptics

# Define required parameters
g = 1.0
Δ = -0.1
α = 4.0

# Define bases for cavity (Fock) and atom (Spin-1/2)
bc = FockBasis(40)
ba = SpinBasis(1//2)

# Construct operators
a = destroy(bc) ⊗ one(ba)
σ⁻ = one(bc) ⊗ sigmam(ba)

# Construct Hamiltonian
H = Δ*dagger(a)*a + g*(dagger(a)*σ⁻ + a*dagger(σ⁻))
```

```
# Define initial state
ψ₀ = coherentstate(bc, α) ⊗ spindown(ba)

# Define list of time steps
T = [0:0.01:35;]

# Evolve in time according to Schrödinger's equation
tout, ψₜ = timeevolution.schroedinger(T, ψ₀, H)

# Calculate atomic excitation
excitation = expect(dagger(σ⁻)*σ⁻, ψₜ)
```

Code sample 1: *Jaynes-Cummings model*.

The resulting atomic dynamics is depicted in Fig. 1, where we show the energy stored in the atom as function of time. We clearly see the well-known collapse and revival of coherent oscillations of the energy between atom and the cavity [12]. This is a well studied and numerically nontrivial phenomenon for which no simple analytic solution exists.

## 2. Framework Design

QuantumOptics.jl's design, and especially its interface, closely resembles QuTiP's successful architecture and shows a lot of similarities to the QO Toolbox, yet features a few fundamental differences. First and foremost, QuantumOptics.jl distinguishes itself in the way it treats quantum objects, such as states, operators and super-operators. In QuTiP, quantum objects are more or less equal to their numerical coefficients with regards to a chosen basis. This is a practical and reasonable approach, as, in the end, the aim is to perform numerical calculations with these objects. However, in order to faithfully represent an abstract state in a Hilbert space, one has to keep track of the choice of basis that was made as well. Thus, in QuantumOptics.jl we explicitly track the basis for every quantum object (see Fig. 2a). From the relation

$$|\Psi\rangle = \sum_i \langle u_i|\Psi\rangle |u_i\rangle = \sum_i \Psi_i |u_i\rangle \quad (3)$$

it is evident that a state is defined completely by specifying the coefficients $\Psi_i$ and a basis $\{|u_i\rangle\}$. For operators, however, it makes sense to allow for more than one basis, since an operator is, in general, a mapping from one Hilbert space into another. Therefore, it is associated with two bases – one for the domain and another one for the co-domain. The following equation formalizes this idea as

$$A = \sum_{ij} \langle u_i| A |v_j\rangle |u_i\rangle \langle v_j| = \sum_{ij} A_{ij} |u_i\rangle \langle v_j|, \quad (4)$$

where we call $\{|u_i\rangle\}$ the left basis and $\{|v_j\rangle\}$ the right basis. The generalization to superoperators is straightforward and culminates in storing four different bases.

Introducing this notion of bases has several advantages. On one hand, it adds an additional layer of safety, as for any operation (e.g. for a multiplication) we check whether or not the bases of the two objects involved are compatible. Without this information about the basis the only thing that can be checked is if the Hilbert space dimensions match. On the other hand, the use of bases arguably improves the code's readability. This is of course a subjective assessment but using a basis as a parameter when creating operators, instead of specifying the Hilbert space's dimension only, leads to more understandable code. Additionally, the possibility of dispatching a function like e.g. `momentum()` on a `PositionBasis` as well as on a `MomentumBasis` and obtaining the correct result in both cases, allows for very elegant coding.

Besides the conceptional differences in the understanding of quantum objects, a more tangible distinction to QuTiP and the QO Toolbox is QuantumOptics.jl's choice of the internal representation of their numerical data. While QuTiP and the QO Toolbox both use sparse matrices as their underlying data structure, QuantumOptics.jl takes a more general approach. It defines an abstract operator interface which is implemented by specialized operator types. At this point, primarily dense and sparse matrices are used but there are additional possibilities, as depicted in Fig. 2b. The existence of different data types for operators is mostly transparent to the user as suitable choices are made automatically. Nevertheless, it is always possible to specify the desired operator type explicitly. Admittedly, the increased complexity that comes from this approach can be an additional burden on the user. However, in our opinion, this disadvantage is far outweighed by the improved versatility (see for example Sec. 4.3) and a huge boost in speed in many cases. A more detailed discussion of this claim is provided in Sec. 5.

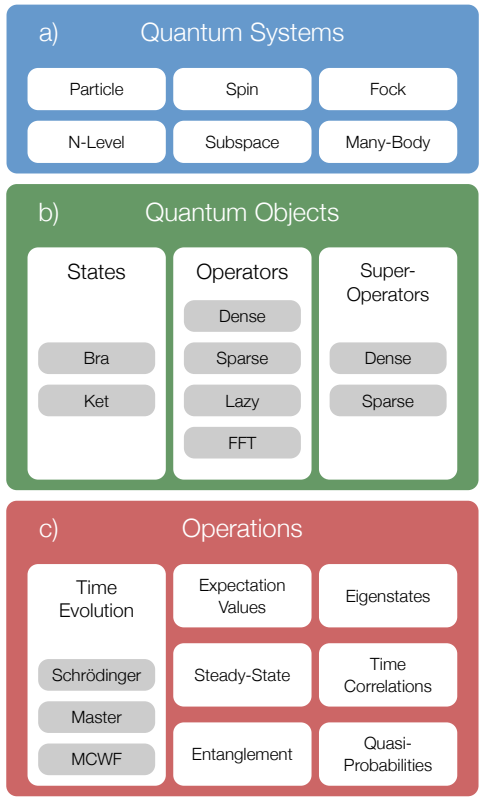

Figure 2: *Illustration of the framework's design.* a) Quantum systems provide functions that allow for an effortless construction of typical quantum objects. b) Quantum objects, i.e. states, operators and superoperators, constitute the fundamental building blocks of QuantumOptics.jl. They are defined as abstract interfaces which are then realized by several specialized types. This makes it possible to choose an implementation that is most favorable for the investigated problem. c) Finally, these quantum objects can be used as an input for various time evolutions as well as other operations.

## 3. Development Philosophy

To ensure the quality as well as the usability of our code we adhere to a certain set of self imposed rules:

*Open source* - Access to the underlying code is a fundamental necessity in any scientific endeavor. This is why our code is open source. Additionally, it can be modified since it is published under the MIT license.

*Open development model* - Framework development takes place transparently using the convenient GitHub platform [13]. Anybody who is interested and motivated can join discussions effortlessly and submit patches. Every single patch is reviewed by at least one person besides its author to ensure high code quality.

*Extensive testing* - Rigorous testing is a core requirement for our code. It enables us to perform restructuring and redesigning while being confident that our code will remain functional. Every function that is part of the public interface is unit-tested. The test suite is run against every single change before it is incorporated into the framework ensuring that even the newest features can be used reliably. Additional high-level tests compare the numerical results against known analytical solutions.

*Documentation* - From a user's point of view undocumented code is equal to nonexisting code. Thus, every function of the public interface is documented directly in the code via docstrings, which can be accessed easily from the command line. High-level documentation [14] can be found on our website [15], which not only references these docstrings but also provides various examples that cover a wide variety of quantum mechanical systems and can be used as a convenient starting point for more specialized investigations. Every single code snippet in the documentation is executed during the build process to guarantee that it is functional and always up to date.

*Benchmarking* - An extensive benchmark suite allows us to detect and therefore avoid any speed regressions [16]. It also includes benchmarks for other quantum simulation frameworks, at the moment for QuTiP and the QO Toolbox, which can be used to identify areas that should be optimized in our own code. A few selected examples are presented in Sec. 5.

## 4. Examples

In this section we demonstrate the versatility of our framework by simulating a few well-known quantum systems using models and techniques of increasing complexity. As is already discernible from code sample 1, each script roughly follows a simple scheme:

1. Define the required physical parameters (frequencies, decay rates, etc.) as numerical constants.
2. Specify the bases of the respective Hilbert spaces (Fig. 2a).
3. Construct the corresponding operators, the Hamiltonian, jump operators and states (Fig. 2b).

4. Use operations such as a time evolution, expectation values, etc. to obtain physical results (Fig. 2c).

*4.1. Lossy Jaynes-Cummings model*

First, let us extend the example of the Jaynes-Cummings model from code sample 1 to an open system. When the cavity mode is coupled to a thermal bath with a mean photon number $n_{\text{th}}$, photons can leak out of the cavity at a rate $(n_{\text{th}} + 1)\kappa$ and enter the cavity at a rate $n_{\text{th}}\kappa$. Similarly, the atom can lose energy via spontaneous emission at a rate $\gamma$ as it interacts with the electromagnetic vacuum field. These dynamics are modeled by a master equation for the system density operator $\rho$ [17],

$$\dot{\rho} = i\left[\rho, H_{\text{JC}}\right] + \mathcal{L}\rho. \quad (5)$$

Here, $\mathcal{L}$ is the Liouvillian which includes the various dissipation channels,

$$\mathcal{L}\rho = (n_{\text{th}} + 1)\kappa\mathcal{D}[a]\rho + n_{\text{th}}\kappa\mathcal{D}\left[a^\dagger\right]\rho + \\ + \gamma\mathcal{D}\left[\sigma^-\right]\rho, \quad (6)$$

where

$$\mathcal{D}[A]\rho = A\rho A^\dagger - \frac{1}{2}\left(A^\dagger A\rho + \rho A^\dagger A\right). \quad (7)$$

The framework is built in such a way that one can easily extend the code from the unitary evolution in code sample 1 to dissipative dynamics. This is shown in code sample 2, where instead of a Schrödinger equation for a pure quantum state we solve the master equation, which inherently requires the use of a density matrix to represent the resulting mixed state.

```
# Decay rates
κ = 0.01
γ = 0.01
n_th = 0.75
R = [(n_th + 1)*κ, n_th*κ, γ]

# Jump operators
J = [a, dagger(a), σ⁻]

# Time evolution according to master equation
tout, ρt = timeevolution.master(T, ψ₀, H, J; rates=R)

# Caclulate atomic excitation
excitation = expect(dagger(σ⁻)*σ⁻, ρt)
```

Code sample 2: *Jaynes-Cummings model including decay (requires code sample 1).*

As one can see in Fig. 3, even though we chose small damping rates, $\kappa, \gamma \ll g$, they already suppress the revival of the atomic excitation.

As an alternative to the master equation one can resort to a stochastic time evolution via the Monte Carlo wave-function (MCWF) method [18]. Since for a single trajectory the state of the system is defined completely by a ket $|\psi\rangle$ rather than a density operator $\rho$ it is easier to simulate. However, this gain comes at the expense of requiring time consuming stochastic averaging. Essentially, the MCWF method evolves the state according to a Schrödinger equation with a non-Hermitian Hamiltonian

$$H_{\text{JC}}^{(\text{nh})} = H_{\text{JC}} - \frac{i}{2}\sum_i r_i J_i^\dagger J_i, \quad (8)$$

with randomly occurring quantum jumps connected to the jump operators $J_i \in \{a, a^\dagger, \sigma^-\}$ and the corresponding rates $r_i \in \{(n_{\text{th}} + 1)\kappa, n_{\text{th}}\kappa, \gamma\}$.

Once again, it is straightforward to implement this time evolution with our framework. To this end, let us extend the Jaynes-Cummings model from code sample 1 and code sample 2 further. In code sample 3 we show how to calculate a single MCWF trajectory.

```
# Calculate single MCWF trajectory
tout, ψt = timeevolution.mcwf(T, ψ₀, H, J; rates=R, seed=2)

excitation = expect(dagger(σ⁻)*σ⁻, ψt)
```

Code sample 3: *Monte Carlo wave-function method for the lossy Jaynes-Cummings model (requires code sample 1 and code sample 2).*

In contrast to the average over infinitely many trajectories (result from the master equation approach), single MCWF trajectories still exhibit a revival in the atomic excitation (see Fig. 3) but with different phase and timing.

*4.2. Time-dependent Jaynes-Cummings model*

Very often the Hamiltonian of a problem contains an explicit time-dependent term, modeling e.g. a controlled change of operating parameters or a pulsed excitation. Let us thus demonstrate how to solve such a time-dependent problem in our framework. Consider the Hamiltonian of the Jaynes-Cummings model. In (2) it is written in a frame rotating at the atomic frequency $\omega_a$, resulting in the term $\Delta a^\dagger a$. To eliminate this term as well we change into a frame

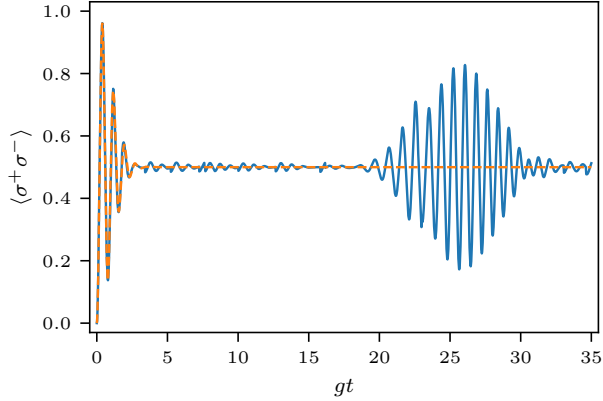

Figure 3: *Jaynes-Cummings model with damping: comparison of full master equation dynamics (orange, dashed) and a single noisy MCWF trajectory (blue, solid). The plot was created with the results from code sample 2 and code sample 3.*

rotating at the detuning $\Delta$. The Hamiltonian then becomes time-dependent,

$$\tilde{H}_{\text{JC}} = g\left(a^\dagger \sigma^- e^{i\Delta t} + a\sigma^+ e^{-i\Delta t}\right). \quad (9)$$

In order to solve the time evolution with a time-dependent Hamiltonian, we need to write a small function that updates the Hamiltonian at every time step and returns the result. In addition, let us also include the decay processes as used in the master equation (see code sample 2). The code required to solve this problem is shown in code sample 4 and we obtain the same results as in code sample 2.

```
# Separate time-dependent terms of H
H₁ = g*dagger(a)*σ⁻
H₂ = dagger(H₁)

# Calculate the Hermitian conjugate for the jump operators
Jdagger = dagger.(J)

function Hₜ(t, ρ) # time-dependent Hamiltonian
  H = exp(1im*Δ*t)*H₁ + exp(-1im*Δ*t)*H₂
  return H, J, Jdagger
end

tout, ρₜ = timeevolution.master_dynamic(T, ψ₀, Hₜ; rates=R)
excitation = expect(dagger(σ⁻)*σ⁻, ρₜ)
```

Code sample 4: *Time evolution of the Jaynes-Cummings model with a time-dependent Hamiltonian (requires code sample 1 and code sample 2).*

### 4.3. Gross-Pitaevskii equation

In addition to the implementation of time-dependent Hamiltonians, the framework also allows for state-dependent effective Hamiltonians. This enables one to, for example, quite easily simulate the Gross-Pitaevskii equation [19]. For a one-dimensional Bose-Einstein condensate (BEC) in free space described by $\psi(x,t)$, the equation reads

$$i\dot{\psi}(x,t) = H_{\text{GPE}}\psi(x,t), \quad (10)$$

where the Hamiltonian is

$$H_{\text{GPE}} = \frac{p^2}{2m} + g|\psi(x,t)|^2. \quad (11)$$

Here, $x$ and $p$ are the position and momentum operators, respectively, and $m$ is the mass. The parameter $g$ governs whether the interaction is attractive ($g < 0$) or repulsive ($g > 0$). The equation has the form of a Schrödinger equation, but with the state-dependent term $g|\psi|^2$ in the Hamiltonian. The implementation of this equation is shown in code sample 5. In our example, the condensate is in a superposition of two counter-propagating wave-packets initially. These collide after some time as depicted in Fig. 4, where the probability density of the BEC $|\psi(x,t)|^2$ is plotted as a function of space and time.

```
using QuantumOptics

x_min = -10
x_max = 10
x_steps = 300
dx = (x_max - x_min)/x_steps
m = 1
x₀ = 2π
g = -3.33

b_x = PositionBasis(x_min, x_max, x_steps)
b_p = MomentumBasis(b_x)

Tpx = transform(b_p, b_x)
Txp = transform(b_x, b_p)

p = momentum(b_p)
Hkin = LazyProduct(Txp, p^2/2m, Tpx)
Hψ = diagonaloperator(b_x, Ket(b_x).data) # ∝ |ψ|^2
H₀ = LazySum(Hkin, Hψ)

function H(t, ψ) # Update state-dependent term in H
  Hψ.data.nzval .= g/dx*abs2.(ψ.data)
  return H₀
end

p₀ = 2
σ = 1.5
ψ₁ = gaussianstate(b_x, -x₀, p₀, σ)
ψ₂ = gaussianstate(b_x, x₀, -p₀, σ)
ψ₀ = normalize(ψ₁ + ψ₂)
T = [0:0.01:6;]
tout, ψₜ = timeevolution.schroedinger_dynamic(T, ψ₀, H)

density = [abs2(ψ.data[j]) for ψ=ψₜ, j=1:x_steps]
```

Code sample 5: *Gross-Pitaevskii equation.*

Besides the state-dependent Hamiltonian this ex-ample showcases a few additional features which are unique to QuantumOptics.jl. For performance reasons it is advantageous to continuously switch between position and momentum space since some operators are sparse in one but not in the other basis. The transformation operators `Txp` and `Tpx` are of a special type, which adheres to the general operator interface, and implicitly performs fast Fourier transformations (FFTs) when multiplied with a state. Since these FFT operators cannot be combined with sparse or dense operators in a meaningful way without losing their advantage, QuantumOptics.jl provides the concept of lazy operators which can be used to delay evaluation until the operator has been applied to a state. For example, instead of adding two operators first and multiplying them with a state afterwards, lazy operators make it possible to first separately multiply the two operators with the state and then sum up the two results.

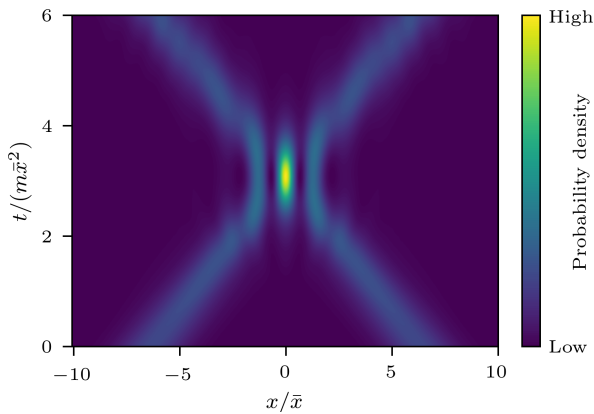

Figure 4: *Nonlinear Schroedinger equation: collision of two soliton like wave-packets*. The plot was created with the results from code sample 5 and $\bar{x}$ is a characteristic length scale.

*4.4. Semi-classical model of cavity cooling*

When a physical system has too many degrees of freedom for an efficient full quantum solution, it is often convenient to approximate a part of the dynamics classically, while other quantum degrees of freedom are kept. Our framework easily allows for implementations of simulating such *semi-classical* dynamics.

Let us demonstrate this at the example of a two-level atom moving in the field of a coherently driven cavity mode along the cavity axis. The cavity has a mode function $\cos(kx)$, where $k$ is the cavity mode wave number and $x$ represents the position of the atom. The full quantum system can be described by the Hamiltonian [20]

$$H_{\text{cooling}} = -\Delta_c a^\dagger a - \Delta_a \sigma^+ \sigma^- + \eta \left(a^\dagger + a\right) + \\ + g\cos(kx)\left(a^\dagger \sigma^- + a\sigma^+\right) + \frac{p^2}{2m}. \quad (12)$$

Here, $\Delta_i = \omega_p - \omega_i$ is the detuning from the pump laser with frequency $\omega_p$ and amplitude $\eta$. The coupling strength between the atom and the cavity is $g$ and $p$ is the momentum of the atom. The atom is subject to spontaneous emission with a rate $\gamma$ and the cavity is damped with a rate $\kappa$, which we include by the Liouvillian

$$\mathcal{L}[\rho] = \kappa \mathcal{D}[a]\rho + \gamma \mathcal{D}\left[\sigma^-\right]\rho. \quad (13)$$

Describing the field, atomic motion and internal atomic dynamics quantum mechanically creates a very large Hilbert space for realistic photon numbers and velocities. However, if the atom has a kinetic energy that is far above the recoil limit, it is well justified to approximate the atomic motion by classical Newtonian mechanics and variables, i.e. $x$ and $p$ are merely numbers instead of operators. This is the case for cavity cooling. The force exerted on the atom is then

$$\dot{p} = -\partial_x \langle H_{\text{cooling}} \rangle = \\ = 2gk\sin(kx)\text{Re}\left\{\langle a^\dagger \sigma^- \rangle\right\}, \quad (14)$$

and the velocity is given by $\dot{x} = p/m$. Tuning the cavity to a frequency lower than the atomic transition frequency, the atom loses kinetic energy upon absorbing a photon from the cavity. It therefore experiences friction and its motion is cooled.

The above model can be simulated in a straightforward fashion using the implemented semi-classical functions, as demonstrated in code sample 6. The resulting cooling process is depicted in Fig. 5.

```julia
using QuantumOptics

κ = 1.0
η = 1.0
g = 0.5
γ = 2.0
Δc = 0.0
Δa = -1.0
m = 3.33
```

```
bc = FockBasis(16)
ba = SpinBasis(1//2)
a = destroy(bc) ⊗ one(ba)
σ⁻ = one(bc) ⊗ sigmam(ba)

Hc = -Δc*dagger(a)*a + η*(a + dagger(a))
Hat = -Δa*dagger(σ⁻)*σ⁻
H₀ = Hc + Hat # Position-independent part
Hₓ = g*(a*dagger(σ⁻) + dagger(a)*σ⁻) # ∝ cos(x)

rates = [κ, γ]
J = [a, σ⁻]
Jdagger = dagger.(J)

function f_q(t, ψ, u) # Quantum part
  x = real(u[1])
  return H₀ + Hₓ*cos(x), J, Jdagger, rates
end

atsm = dagger(a)*σ⁻ # Compute a priori for efficiency
function f_cl(t, ψ, u, du) # Classical part
  x, p = real(u)
  du[1] = p/m
  du[2] = 2g*sin(x)*real(expect(atsm, ψ))
end

x₀ = -2π
p₀ = 2m
u₀ = Complex128[x₀, p₀]
ψ₀ = fockstate(bc, 0) ⊗ spindown(ba)
ψsc = semiclassical.State(ψ₀, u₀)

T = [0:0.1:100;]
tout, ρₜ = semiclassical.master_dynamic(T, ψsc, f_q, f_cl)

x = [r.classical[1] for r=ρₜ]
p = [r.classical[2] for r=ρₜ]
n = expect(dagger(a)*a, ρₜ)
```

Code sample 6: *Semi-classical model of cavity cooling.*

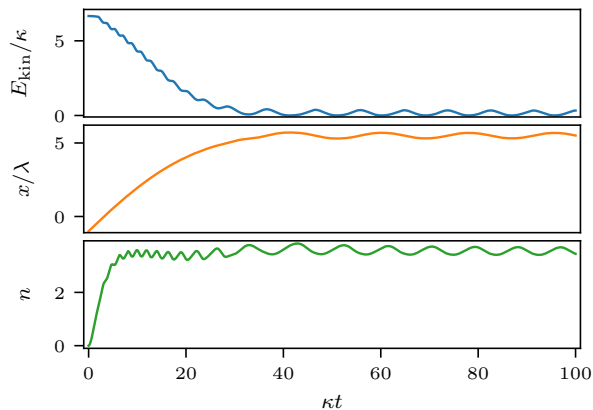

Figure 5: *Cavity cooling of a two-level atom.* The plot was created from the results in code sample 6. One can see the cooling process in terms of the decrease of the kinetic energy $E_{\text{kin}} = p^2/2m$ and the localization in $x$ (where $\lambda = 2\pi/k$ is the cavity wavelength). Additionally, the cavity photon number $n$ saturates.

## 5. Performance

Besides the correctness of the numerical results, providing an adequate performance is one of our main goals. An extensive benchmark-suite [16] allows us to detect and therefore avoid speed regressions from one version to the next. Additionally, since these benchmarks contain respective tests for QuTiP and the QO Toolbox as well, it helps us to identify areas in our code that should be optimized further.

In principle, there are two intrinsic aspects that should give us an advantage in terms of speed. The first one is the fact that the internal layout of operators can be chosen according to the investigated problem. This means that one can choose to work with dense or sparse matrices or even more specialized operators. Note, that at the moment the time-evolution methods in our framework require density operators to be represented as dense operators. There is no fundamental reason for this limitation and a future implementation of time evolutions with sparse density operators is part of our roadmap. The positive effect of having a choice between dense and sparse operators can be seen best in the benchmarks in Fig. 6 which compare the different frameworks by performing a time-evolution according to a master equation for three different physical examples. Here, the cavity example exhibits a sparse Hamiltonian and a dense density operator, the Jaynes-Cummings model features a sparse Hamiltonian and a relatively sparse density operator, while the particle example demonstrates a completely dense Hamiltonian as well as a dense density operator. The more sparseness the whole system entails the closer the benchmark results are to each other. Contrastingly, for the low sparsity case huge speed improvements are observable.

The second aspect is Julia's natural speed advantage in comparison to Python and Matlab. While code written in Julia can achieve the speed of C or Fortran, in interpreted languages the philosophy is to rewrite speed-critical parts in a fast compiled language. However, this comes at the cost of flexibility, which manifests itself, e.g. when simulating time- or especially state-dependent problems. The simplest way for the user would be to define an arbitrary function directly within Python or Matlab. However, this function is most likely critical for performance, which means it should be compiled. QuTiP uses Cython [6] to achieve this and, as can be seen in the time-dependent benchmarks in Fig. 7, quite successfully so. This again comes at the cost of accessibility, though.

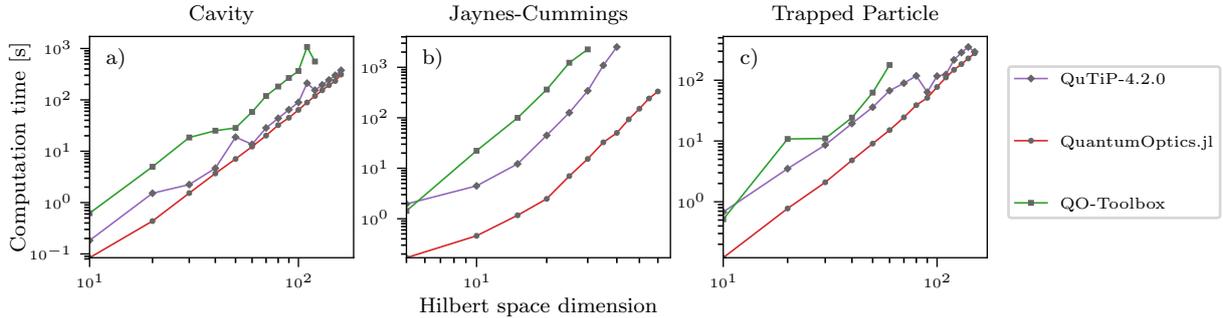

Figure 6: *Benchmarks measuring the time elapsed when performing a time-evolution according to a master equation.* Three different systems exhibiting different sparseness properties are investigated: a) a pumped cavity with photon decay (sparse Hamiltonian and dense density operator), b) the Jaynes-Cummings model including particle as well as cavity decay (sparse Hamiltonian and sparse density operator), c) a particle trapped in a harmonic potential (dense Hamiltonian and dense density operator). Depending on the sparseness of the system, QuantumOptics.jl's flexible operator types can lead to considerable speed-ups compared to the purely sparse matrix approach in QuTiP and the QO Toolbox.

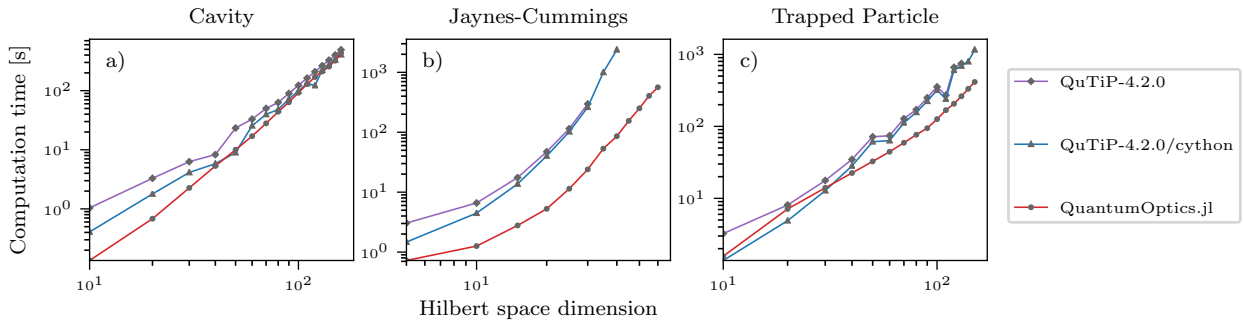

Figure 7: *Benchmarks measuring the time elapsed when performing a time-evolution according to a master equation with time-dependent Hamiltonians.* The same systems as in Fig. 6 are depicted in the same order. QuTiP provides two different implementations - a Cython and a pure Python based approach. Both, the compilation time of Julia and the Cython approach are not included in the above time measurements. Note, though, that in general the GCC compiler used by Cython gives it a certain advantage over Julia (see Sec. 6)

While in this publication we focus on the speed of performing various time-evolutions, our benchmarks cover a greater variety of calculations. This is especially important when investigating more involved examples, e.g. when treating a system semiclassically as in the example in Sec. 4.4, where in every step of the time evolution expectation values have to be calculated.

Finally, let us provide a few more details on how these benchmarks were carried out: the data shown in this paper was obtained on a single core of an Intel(R) Core(TM) i7-5960X CPU running at 3.00GHz under Linux and Julias's compile time is neglected in the results as we focus on the speed of the actual calculations. Furthermore, we tested QuTiP and QuantumOptics.jl on other hardware and different operating systems obtaining qualitatively similar results.

## 6. Disadvantages

While QuantumOptics.jl relies on Julia and embraces many of its modern features, we have to acknowledge the fact that Julia is still a very young language and not yet completely stable. The very same is true for the framework itself, which will continue to grow and future changes to the inter-face might very well be in its path. This, however, implies that code written for the current release of QuantumOptics.jl, namely v0.4.1, might have to be adopted to its future versions, if they include changes in the interface.

Another consequence of the framework being very young is the fact that QuantumOptics.jl is clearly not as feature-rich as other well-established frameworks such as QuTiP. We are confident, though, that over time the framework will continue to grow and eventually reach a comparable versatility.

A clear disadvantage when comparing Julia to Cython is the compilation time. We want to point out that Julia uses the noticeably slower LLVM compiler, while Cython relies on GCC making it advantageous at short running times. When going to longer times, however, this difference in compilation time becomes a negligible constant offset.

Another noteworthy issue lies with the Julia language itself. As of the writing of this manuscript, Julia has quite a large memory footprint. Hence, so does our framework in its current form. This may be an issue especially for users running many simultaneous instances of Julia (e.g. when performing calculations on a cluster or server). We hope that this issue will be addressed in the future of Julia's development.

## 7. Conclusions & Outlook

We have presented a new computational framework for the efficient numerical investigation of open quantum systems, demonstrated its capabilities and highlighted its performance.

In its current version (v0.4.1), QuantumOptics.jl is a very young framework and, as mentioned above, still under active development. We strongly encourage community contributions in the form of additions to the framework itself or even separate extensions based on QuantumOptics.jl, like our own Correlation Expansion Package [21], which allows for simulating larger systems by specifying which quantum correlations should be included or neglected, respectively, or the CollectiveSpins library [22], providing more specialized building blocks for the investigation of dipole-dipole coupled two-level systems in various approximations. Planned short term improvements include the addition of stochastic Schrödinger and master equations as well as adopting DifferentialEquations.jl [23] for the time evolution functions.

## 8. Acknowledgements

We acknowledge financial support by the Austrian Science Fund (FWF) through projects SFB FoQus F4006-N13 (S.K. & H.R), the DK-ALM: W1259-N27 (D.P.) and P29318-N27 (L.O.). Graphs were done with the open source plotting library Matplotlib [24].